\documentclass[aps,preprint,amsfonts,superscriptaddress,byrevtex]{revtex4}
\usepackage{epsfig}
\usepackage{graphicx}
\begin{document}
\title{Stability of racemic and chiral steady states in open and closed chemical systems}
\author{Josep M. Rib\'{o}}
\affiliation{Departament de Qu\'{i}mica Org\`{a}nica, Universitat de
Barcelona, c. Mart\'{i} i Franqu\`{e}s 1, Barcelona, Spain}
\author{David Hochberg}
\email{hochbergd@inta.es} \homepage[]{http://www.cab.inta.es}
\affiliation{Centro de Astrobiolog\'{\i}a (CSIC-INTA), Ctra. Ajalvir
Km. 4, 28850 Torrej\'{o}n de Ardoz, Madrid, Spain}


\begin{abstract}
The stability properties of models of spontaneous mirror symmetry
breaking in chemistry are characterized algebraically. The models
considered here all derive either from the Frank model or from
autocatalysis with limited enantioselectivity. Emphasis is given to
identifying the critical parameter controlling the chiral symmetry
breaking transition from racemic to chiral steady-state solutions.
This parameter is identified in each case, and the constraints on
the chemical rate constants determined from dynamic stability are
derived.
\end{abstract}

\maketitle

\section{\label{sec:intro} Introduction}

Ever since the Soai reaction, first published some years ago
\cite{Soai}, there has been an ever increasing number of researchers
inspired by this important landmark experiment, who have been
searching for other reactions capable of exhibiting the spontaneous
emergence of chiral asymmetry in closed systems. At present, there
is a gradual and mounting experimental evidence in favor of this
\cite{Viedma,Martynova,Noorduin}, showing that absolute asymmetric
synthesis \cite{Caglioti} is possible for autocatalytic reaction
networks in closed mass reacting systems and even suggests that
chiral amplification could occur in chemical equilibrium scenarios
\cite{Crusats}. On the theoretical side, kinetic schemes which are
extensions of the elemental Frank model \cite{Frank}, are able to
reproduce the main features of the mirror symmetry breaking behavior
of the Soai reaction \cite{Islas}.

The key ingredients of theoretical models of mirror-symmetry
breaking processes in chemistry \cite{all} include reactions in
which the chiral products serve as catalysts to produce more of
themselves while inhibiting the production of their enantiomer or
mirror-image counterparts. The two basic models we analyze here
differ in the way this inhibition is produced. Frank's original
model \cite{Frank,Hochstim,all,Mason,Mason2}, involves the
autocatalysis of the two enantiomers, denoted herewith as L and D,
and mutual inhibition or antagonistic effects between the two chiral
species. This mutual inhibition occurs through the formation of
heterodimers that are removed from the reacting system. In the case
of limited enantioselectivity \cite{Avet}, the required antagonism
arises through the recycling of an enantiomeric pair of monomers
back to a single chiral monomer and a prochiral substrate molecule.

The purpose of this paper is to determine the stability of the
steady states for various reaction schemes that have been proposed
as models for mirror symmetry breaking in chemistry. This requires
knowledge of only the static solutions of the kinetic equations.
Then, differential equations for the arbitrary time-dependent
fluctuations about these final static solutions are straightforward
to derive to first order in the fluctuations. These time dependent
fluctuation equations are expressed in matrix form, and the
eigenvalues of the corresponding matrix, evaluated on the static
solutions obtained previously, signal unambiguously the stability
(or instability) of the racemic or chiral solution under study. Here
we are interested in understanding the chemical factors controlling
the symmetry breaking transition from a racemic (equal proportions
of the left and right-handed molecules) to chiral states, the latter
characterized by unequal proportions of the two enantiomers. For
each model considered, we identify the critical parameter that
controls the transition from racemic to chiral states. These
dimensionless parameters are expressed as simple ratios of the
chemical reaction rates. When the control parameter falls below a
critical value, then chiral amplification results. In the latter
case, the racemic state is unstable, and an initial small
perturbation is sufficient to tip the system over into one of its
two equally likely stable chiral states. Exactly equal proportions
of the two chiral enantiomers never occur in practice, and this
inevitable chiral imbalance yields an initial statistical
enantiomeric excess \cite{Mills}. Moreoever, in imperfectly mixed
spatially extended systems, diffusion-limited noise is intrinsic to
the reacting and diffusing system itself and is sufficient to drive
the symmetry breaking \cite{Lente,HZ,HZ2}. In either case, the noise
need not be put in \textit{a-posteriori}.

First the study of open systems at constant concentration of the
achiral reactants is carried out. Open systems provide a clear
insight to the chemical conditions necessary to achieve spontaneous
mirror symmetry breaking; as we will show, the critical parameter
for the bifurcation behavior in a closed system is concentration
dependent. The calculation of the eigenvalues in the closed system
case are generally more difficult to obtain and in many cases, they
cannot be expressed in a manageable analytic form. The reaction
schemes analyzed below correspond in the most part to
\textit{reversible} chemical reactions, in contrast to the
irreversible Soai reaction. The results obtained here should be of
practical value to aid in the search for new reactions exhibiting
chiral symmetry breaking which might eventually provide valuable
insight to the much more difficult problem of the origin of
biological chirality \cite{Palyi}. The importance of cyclic
catalytic reactions as conditions for life is elegantly argued in
\cite{Eigen}.

This paper is organized as follows. After introducing the general
reaction scheme in Sec \ref{sec:mainmodel}, we consider open flow
systems, and use the constancy of the prochiral substrate to cast
the kinetic equations in dimensionless form in Sec
\ref{sec:opensys}. This has the advantage of reducing the number of
parameters and allows for all results to be expressed in terms of
\textit{ratios} of the rate constants. This is a most welcome
feature when calculating solutions and the corresponding eigenvalues
of the stability matrix. The stability properties of a sequence of
Frank-type models followed by models with limited enantioselectivity
are presented in Sec \ref{sec:stability}. The symmetry breaking in
all these models yields the classic bifurcation diagram which we
illustrate for the limited enantioselectivity model.  For closed
systems, a different rescaling of time and the concentrations is
needed to obtain dimensionless kinetic equations. The constant mass
constraint makes the algebraic analysis in this case much more
involved, and a few particularizations of the general reaction
scheme are treated in Sec \ref{sec:closedsystems}. Conclusions are
drawn in Sec \ref{sec:disc}.

\section{\label{sec:mainmodel} The general kinetic model}

The general system we will study for spontaneous mirror symmetry
breaking is that defined by Eqs.(\ref{decay}-\ref{heterodim}) below
and it implies the following four pairs of reactions: a straight
non-catalyzed reaction Eq.(\ref{decay}), enantioselective
autocatalysis Eq.(\ref{autoLD2}), and a non-enantioselective
autocatalysis Eq.(\ref{nonenantiosel}), where A is a prochiral
starting product, and L and D are the two enantiomers of the chiral
product. We also assume reversible homo- and heterodimerization
steps in Eqs.(\ref{homodim},\ref{heterodim}); $\rm L_2,D_2$
represent the two chiral enantiomeric homodimers and LD is the
diastereomeric, achiral heterodimer. The $k_i$ denote the reaction
rate constants.

\noindent Production of chiral compound:
\begin{equation}\label{decay}
\textrm{A} \stackrel{k_1}{\rightleftharpoons \atop{\small k_{-1}}}
\textrm{L}, \qquad \textrm{A} \stackrel{k_1}{\rightleftharpoons
\atop{\small k_{-1}}} \textrm{D}.
\end{equation}

\noindent Autocatalytic production:
\begin{equation}\label{autoLD2}
\textrm{L} + \textrm{A} \stackrel{k_2}{\rightleftharpoons
\atop{\small k_{-2}}}\textrm{L} + \textrm{L}, \qquad \textrm{D} +
\textrm{A} \stackrel{k_2} {\rightleftharpoons \atop{\small k_{-2}}}
\textrm{D} + \textrm{D}.
\end{equation}

\noindent  Limited enantioselectivity:
\begin{equation}\label{nonenantiosel}
\textrm{L} + \textrm{A} \stackrel{k_3}{\rightleftharpoons
\atop{\small k_{-3}}} \textrm{L} + \textrm{D}, \qquad \textrm{D} +
\textrm{A} \stackrel{k_3}{\rightleftharpoons \atop{\small k_{-3}}}
\textrm{L} + \textrm{D}.
\end{equation}

\noindent  Homo-dimerizations:
\begin{equation}\label{homodim}
\textrm{L} + \textrm{L} \stackrel{k_4}{\rightleftharpoons
\atop{\small k_{-4}}} \textrm{L}_2, \qquad \textrm{D} + \textrm{D}
\stackrel{k_4}{\rightleftharpoons \atop{\small k_{-4}}}
\textrm{D}_2.
\end{equation}

\noindent  Hetero-dimerization:
\begin{equation}\label{heterodim}
\textrm{L} + \textrm{D} \stackrel{k_5}{\rightleftharpoons
\atop{\small k_{-5}}} \textrm{LD}.
\end{equation}
We assume for all reaction steps the feasibility of the reverse
reaction, and that the reaction rates are the same for each pair of
enantiomeric reactions Eqs.(\ref{decay}-\ref{homodim}). Notice that
this would not be the case in the presence of an external chiral
polarization field, or for an internal bias of non-chemical origin,
such as the weak nuclear force \cite{Quack}, for then $k_{\pm i}^L
\neq k_{\pm i}^D$. This reaction scheme leads to the following
differential equations for the concentrations in the mean field
limit:
\begin{eqnarray}\label{eqnaa}
\frac{d}{d t}[L] &=& k_1[A] + (k_2[A]-k_{-1})[L]-k_{-2}[L]^2 -
k_{-3}[L][D] + k_3[A][D] -2k_4[L]^2 + 2k_{-4}[L_2]\nonumber \\
&-& k_5[L][D] + k_{-5}[LD], \nonumber \\
\label{eqnbb} \frac{d}{d t}[D] &=& k_1[A] + (k_2[A]-k_{-1})[D]
-k_{-2}[D]^2 -k_{-3}[D][L] + k_3[A][L] -2k_4[D]^2 +
2k_{-4}[D_2]\nonumber \\
&-& k_5[D][L] + k_{-5}[LD], \nonumber \\
\frac{d}{dt}[A] &=& -2k_1[A]-(k_2[A]+k_3[A] -k_{-1})([L]+[D])+
k_{-2}([L]^2 + [D]^2) +2k_{-3}[L][D],\nonumber \\
\frac{d}{d t}[L_2] &=& k_4[L]^2 - k_{-4}[L_2], \nonumber \\
\frac{d}{d t}[D_2] &=& k_4[D]^2 - k_{-4}[D_2], \nonumber \\
\frac{d}{d t}[LD] &=& k_5[L][D] - k_{-5}[LD]. \nonumber
\end{eqnarray}
The other important feature that relates our work to experimental
chemical systems is the inclusion of both the forward $k_i$, and the
reverse $k_{-i}$ reaction rates, despite the fact that for some rate
constants one has $k_i
>> k_{-i}$. This inclusion is necessary
because these rates determine the thermodynamic conditions, i.e., the principle
of detailed balance, microscopic reversibility; see for example the
final paragraph of Section D

\section{\label{sec:opensys} Open flow systems}

The other experimental condition that comes into play is the
distinction between open and closed systems. Open flow is
traditionally invoked for maintaining the non-equilibrium state. In
these models, this is achieved by providing the system with a
continuous input flow of achiral precursers A and an output flow of
certain products (such as the heterodimers P in the classic Frank
model). Openess for chemical systems is cannot be applied to recent
experimental reports on the chemical bench (i.e., the Soai
reaction). Closed on the other hand, refers to systems of constant
mass where only energy can be exchanged with the exterior: the start
from an initial state far from equilibrium to the final composition
state is a real representation of a chemical reaction carried out in
the laboratory. Here we consider reactions that take place in an
open system. Open here specifically means that the system can
exchange the substrate A with an external source, and thus we take
the concentration $[A]$ as a constant. This is also in keeping with
the original formulation of the Frank model \cite{Frank,Mason}. So
we impose this condition and work out the consequences for a
sequence of models starting from the Frank model itself. It proves
convenient to use this constant concentration to define a simple
transformation to dimensionless rates, time and concentrations. We
take $\tau = (k_2[A]-k_{-1})t$ for the time parameter and $[\tilde
L] = \frac{k_{-3} + k_5}{k_2[A]-k_{-1}}[L]$, etc. for the
dimensionless concentrations. This allows us to express the above
rate equations in pure dimensionless form:
\begin{eqnarray}\label{eqntilL}
\frac{d}{d \tau}[\tilde L] &=& u + [\tilde L] - g[\tilde L]^2
-[\tilde L][\tilde D]
+w[\tilde D] + 2p[\tilde L_2] + r[\tilde{LD}], \nonumber \\
\label{eqntilD} \frac{d}{d \tau}[\tilde D] &=& u + [\tilde D] -
g[\tilde D]^2 -[\tilde D][\tilde L] +w[\tilde L] + 2p[\tilde D_2] +
r[\tilde{LD}],
\nonumber \\
\frac{d}{d \tau}[\tilde L_2] &=& \frac{q}{2}[\tilde L]^2 - p[\tilde L_2], \nonumber \\
\frac{d}{d \tau}[\tilde D_2] &=& \frac{q}{2}[\tilde D]^2 - p[\tilde D_2], \nonumber \\
\frac{d}{d \tau}[\tilde{LD}] &=& s[\tilde L][\tilde D] -
r[\tilde{LD}]. \nonumber
\end{eqnarray}
The dimensionless parameters appearing here are:
\begin{eqnarray}\label{openparams1}
u &=& \frac{k_1[A](k_{-3}+k_5)}{(k_2[A]-k_{-1})^2},\,\, g
=\frac{k_{-2}+ 2k_4}{k_{-3}+k_5}, \,\,
w=\frac{k_3[A]}{(k_2[A]-k_{-1})}, \,\, p =
\frac{k_{-4}}{(k_2[A]-k_{-1})}, \\\label{openparams2} r &=&
\frac{k_{-5}}{(k_2[A]-k_{-1})}, \,\,q = \frac{2k_{4}}{(k_{-3}+
k_5)}, \,\, {\rm and} \,\, s= \frac{k_{5}}{(k_{-3}+ k_5)}.
\end{eqnarray}
Note we have succeeded in reducing the number of parameters from ten
to seven. These latter seven parameters fully determine the dynamics
of the kinetic scheme. We will see below that $g$ plays a privileged
role. Depending on whether $g$ is greater or less than a certain
critical value $g_{crit}$, then either the racemic or chiral
solutions will be stable, respectively. Thus we already see the
relative influence that the homo ($k_4$) and heterodimerizations
($k_5$) will have on the final outcome as well as the relative
influence of the \textit{reverse} autocatalytic ($k_{-2}$) and
limited enantioselective steps ($k_{-3}$). From the denominator of
$g$, we see that the chiral antagonism proceeds via two independent
pathways: either through the formation of heterodimers $(k_5)$, or
via the recycling of an enantiomeric pair of monomers back to a
single chiral monomer and the prochiral substrate $(k_{-3})$.

To cast the equations in their final form, we next define sums and
differences of the (dimensionless) monomer and the homodimer
concentrations. Thus, we will put $\chi = [\tilde L] + [\tilde D]$,
$\chi_2 = [\tilde L_2] + [\tilde D_2]$, $y = [\tilde L] - [\tilde
D]$, $y_2 = [\tilde L_2] - [\tilde D_2]$, and $P = [\tilde{LD}]$.
This then yields the following:
\begin{eqnarray}\label{chi}
\frac{d}{d \tau}\chi &=& 2u + \chi -\frac{1}{2}(g+1)\chi^2
-\frac{1}{2}(g-1)y^2 +w\chi + 2p\chi_2 +2r P,\\\label{y} \frac{d}{d
\tau}y &=& y(1-w -g\chi) + 2p\,y_2,\\\label{chi2}
\frac{d}{d \tau}\chi_2 &=& \frac{q}{4}(\chi^2 + y^2) -p\,\chi_2, \\
\label{y2} \frac{d}{d \tau}y_2  &=& \frac{q}{2}\,\chi y- p\,y_2,
\\ \label{P}
\frac{d}{d \tau} P &=&  \frac{s}{4}(\chi^2-y^2) - r P.
\end{eqnarray}
The fixed points of these equations correspond to the \textit{final}
asymptotic (as $\tau \rightarrow \infty$) solutions of the kinetic
model. Substituting $\chi = \chi^* + \delta \chi(t), y = y^* +
\delta y(t)$, etc. into the set of Eqs.(\ref{chi}-\ref{P}) where
$\chi^*, y^*,...,$ denotes a fixed point solution, we obtain
differential equations for the arbitrary perturbations $\delta
\chi(t), \delta y(t), ...,$ about the fixed point. Then, the
following Jacobian matrix $M_{open}$ governs the time dependence of
these perturbations to first order $O(\delta)$ in the fluctuations:
\begin{equation}\label{Mopen}
M_{open} = \left(
  \begin{array}{ccccc}
    1-(g+1)\chi +w & (1-g)y & 2p & 0 & 2r \\
    -gy & 1-w-g\chi & 0 & 2p & 0 \\
    \frac{q}{2}\chi & \frac{q}{2}y & -p & 0 & 0 \\
     \frac{q}{2} y &  \frac{q}{2}\chi & 0 & -p & 0 \\
    \frac{s}{2}\chi & -\frac{s}{2}y & 0 & 0 & -r \\
  \end{array}
\right).
\end{equation}
This matrix is to be evaluated on any one of the static fixed-point
solutions $\chi^*,y^*,\chi_2^*,y_2^*,P^*$ of the equation set
Eqs.(\ref{chi}-\ref{P}). The associated eigenvalues indicate the
stability of the specific static solution on which $M_{open}$ is
evaluated.

The eigenvalues of this Jacobian are expressed in terms of
dimensionless ratios of specific chemical reaction rates, thus, the
positivity or negativity of the individual eigenvalues can be
ascertained straightforwardly in terms of the specific chemical
parameters. This provides another clear connection with the
experimental conditions: the relative rates of certain reactions
will determine whether the final outcome is a chiral or racemic
solution. Of course, here we treat the $k$'s as variables in order
to analyze all possibilities of chemical $k$ rates. Thus all cases
for different values and relationships between rate constants are
taken into account in our analysis: this allows us to characterize
the stable final states for reversible and quasi reversible, and
also irreversible, synthetic chemical reactions formed by the
reaction networks analyzed. These are the most widely studied as
complex chemical systems capable of leading to spontaneous symmetry
breaking in previous works. In practice, the rate constants can be
determined experimentally in the laboratory, such methods are
explained in detail in textbooks \cite{Benson,Chang}.

\section{\label{sec:stability} Stability properties and critical parameters: open systems}

\subsection{\label{sec:Frank} The Frank model}
First consider the Frank model as determined by the kinetic scheme
Eqs.(\ref{decay},\ref{autoLD2},\ref{heterodim}). See, for example
the definition as given in \cite{Mason,GTV}. The concentration $[A]$
of the achiral substrate is taken as constant. There is a reversible
catalytic production of the monomers. The heterodimers $[LD]$,
formed irreversibly ($k_{-5} = 0$) from the mutual inhibition step,
are to be eliminated from the system as a kind of inactive side
product. The production step Eq.(\ref{decay}) is typically ignored
(we will however consider its effect below). This situation
corresponds then to the reaction steps
Eqs.(\ref{autoLD2},\ref{heterodim}), and we are to solve the
equations Eqs.(\ref{chi},\ref{y}) after setting $u=w=p=r=0$. There
are four fixed points or static solutions:
\begin{eqnarray}\label{Franksolu0}
\begin{array}{lll}
O & \equiv (\chi =0 , y = 0) \\
R & \equiv (\chi = \frac{2}{1 + g} , y = 0) \\
Q_{\pm}& \equiv \left(\chi = \frac{1}{g}  , y = \pm \frac{1}{g}\,
\right).
\end{array}
\end{eqnarray}
$O$ denotes the empty solution, that is, with zero chiral matter,
$R$ is the racemic solution with positive net total chiral matter,
and $Q_{\pm}$ denote the two possible chiral solutions. From
Eq.(\ref{openparams1}) we see that $g = \frac{k_{-2}}{k_5}$ and
$s=1$.

In order to study the stability of the four possible homogeneous
solutions $O, R$ and $Q_{\pm}$, we calculate the eigenvalues of the
$2\times 2$ subblock of the upper left hand corner of the Jacobian
matrix Eq.(\ref{Mopen}) (after first setting $w=0$ there) evaluated
at each one of the above four solutions Eqs. (\ref{Franksolu0}). The
eigenvalues are given by
\begin{eqnarray}\label{FrankeigOu0}
  \lambda_{1,2}(O) &=& (1,1) \\ \label{FrankeigRuO}
  \lambda_{1,2}(R) &=& \big(-1,\frac{1-g}{1+g}\big)\\ \label{FrankeigQu0}
  \lambda_{1,2}(Q_{\pm}) &=& \Big(-1,\frac{-1+g}{g}\Big). \\
  & & \nonumber
\end{eqnarray}
The empty solution $O$ is always unstable. To ensure the stability
in $R$, we need for both eigenvalues to be negative:
$\lambda_{1,2}(R) <0$. While the first eigenvalue $\lambda_{1}(R) =
-1$ is always negative, we have $\lambda_{2}(R) <0$ if and only if
$g>1$. Concerning the stability of the solutions $Q_{\pm}$, these
are always unstable if $g>1$ (since $\lambda_{2}(Q) > 0$) and they
become stable for $0<g<1$ (since now $\lambda_{2}(Q) < 0$). Thus
$g_{crit} =1$ \cite{GTV}. So, the chiral solution is obtained
provided that the rate of heterodimerization exceeds that of the
reverse autocatalytic step. In the chiral broken solution, $Q_{\pm}$
in Eq. (\ref{Franksolu0}), the final enantiomeric excess $|ee| =
|y/\chi| = |\pm\frac{1}{g}/\frac{1}{g}| = 1$ is always equal to
unity in absolute value. These solutions are homochiral. Note that
the irreversible heterodimer formation is mathematically equivalent
to their elimination from the system as a continuous output, see (d)
in \cite{all}. This can be corroborated from
Eqs.(\ref{decay},\ref{autoLD2},\ref{heterodim}). For
\textit{elimination}, then $P=0$ is constant and equal to zero in
the system, and we solve the two equations
Eqs.(\ref{decay},\ref{autoLD2}). Instead, for irreversible
formation, we set $r=0$. But then Eq.(\ref{heterodim}) for $P$ is
not independent from Eqs.(\ref{decay},\ref{autoLD2}): we solve for
$(\chi,y)$ and then deduce $P$. The above steady state solutions
Eq.(\ref{Franksolu0})  and corresponding eigenvalues
Eqs.(\ref{FrankeigOu0}-\ref{FrankeigQu0}) are identical for either
situation.

We now allow for $u>0$ and determine the effect that the direct
monomer production step Eq.(\ref{decay}) has on the solutions, and
their stability properties. The four static solutions are now given
by
\begin{eqnarray}\begin{array}{lll}
U & \equiv (\chi = \frac{1-\sqrt{1+4(1+g)u}}{1+g} , y = 0) \\
R & \equiv (\chi = \frac{1+\sqrt{1+4(1+g)u}}{1+g} , y = 0) \\
Q_{\pm}& \equiv \left(\chi = \frac{1}{g}  , y = \pm
\frac{1}{g}\sqrt{1-\frac{4ug^2}{1-g}}\, \right).
\end{array}
\end{eqnarray}
It is easy to see that these tend to the solutions in
Eq(\ref{Franksolu0}) in the limit as  $u \rightarrow 0$. Here $U$
denotes the \textit{unphysical} solution, since the total
concentration of chiral matter is negative $\chi < 0$. The
eigenvalues are given by
\begin{eqnarray}
  \lambda_{1,2}(U) &=& \big(\sqrt{1+4(1+g)u},\, \frac{1+g\sqrt{1+4(1+g)u}}{1+g} \big) \\
  \lambda_{1,2}(R) &=& \big(-\sqrt{1+4(1+g)u},\,\frac{1-g\sqrt{1+4(1+g)u}}{1+g}\big)\\
  \lambda_{1,2}(Q_{\pm}) &=& \Big(-\frac{1+\sqrt{1+4g(-1+g+4g^2u)}}{2g},\,
    \frac{-1+\sqrt{1+4g(-1+g+4g^2u)}}{2g}                                      \Big). \\
  & & \nonumber
\end{eqnarray}
In the $u \rightarrow 0$ limit these eigenvalues tend those listed
in Eqs.(\ref{FrankeigOu0}-\ref{FrankeigQu0}). Since
$\lambda_{1,2}(U) > 0$, the unphysical solution is always unstable.
Note also that $\lambda_1(R) <0$ and $\lambda_1(Q) <0$ are always
negative whereas $\lambda_2(R) > 0$ and $\lambda_2(Q) < 0$ for $g <
g_{crit}$ where $g_{crit} = \frac{1}{8u}(\sqrt{1+16u}-1)$. Note that
$g_{crit}(u) \leq 1$ for all $u \geq 0$. For small $u$ we can write
$g_{crit} = 1 - 4u$; while for large $u$, $g_{crit} \rightarrow
\frac{1}{2u^{1/2}}$. Thus the monomer production step tends to
racemize the system, lowers the final $ee$ to values strictly less
than unity:
\begin{equation}
ee = \pm\sqrt{1-\frac{4ug^2}{1-g}},
\end{equation}
and drives the value of $g_{crit}$ below one. The monomer production
step thus reduces the range of $g$ for which stable mirror symmetry
breaking can occur, and the chiral solutions are no longer
homochiral.

\subsection{\label{sec:Frankrevhet} Frank with reversible heterodimer formation}

What happens if the heterodimers are \textit{not} removed from the
system, but are allowed to remain evolving dynamically in concert
with the monomers? This situation corresponds to all three reaction
steps Eqs.(\ref{decay},\ref{autoLD2},\ref{heterodim}), and we solve
the three equations Eqs.(\ref{chi},\ref{y},\ref{P}) after setting
$w=p=q=0$ and $s=1$. To keep the algebra manageable, we also will
set $u=0$. There are four static solutions:
\begin{eqnarray}\begin{array}{lll}
O & \equiv (P=0, \chi = 0 , y = 0) \\
R & \equiv (P= \frac{1}{g^2 r},\chi = \frac{2}{g} , y = 0) \\
Q_{\pm}& \equiv \left(P = 0, \chi = \frac{1}{g}  , y = \pm
\frac{1}{g} \right).
\end{array}
\end{eqnarray}
Note that the final heterodimer concentration $P$ is \textit{zero}
in the chiral states $Q_{\pm}$.

In order to study the stability of the four possible homogeneous
solutions $O, R$ and $Q_{\pm}$, we calculate the eigenvalues of the
$3\times 3$ array obtained from Eq.(\ref{Mopen}) after deleting the
3rd and 4rth rows and columns. The eigenvalues corresponding to
these solutions are given by
\begin{eqnarray}
  \lambda_{1,2,3}(O) &=& (1,1,-r) \\
  \lambda_{1,2,3}(R) &=& \big(-1,\,-\frac{2+g(1+r)+\sqrt{4+g^2(-1+r)^2+4g(1+r)}}{2g}, \nonumber \\
  && \frac{-2-g(1+r)+\sqrt{4+g^2(-1+r)^2+4g(1+r)}}{2g} \big)\\
  \lambda_{1,2,3}(Q_{\pm}) &=& \Big(-1,\,-\frac{1+g(-1+r)+\sqrt{1+2g(-1+r)+g^2(1+r)^2}}{2g}, \nonumber \\
   && \frac{-1+g(1-r)+\sqrt{(1+g(-1+r))^2+4g^2r}}{2g}\Big). \\
  & & \nonumber
\end{eqnarray}
As $\lambda_{1,2}(O) >0$, the empty state is always unstable. An
inequality analysis shows that both $\lambda_2(R) <0$ and
$\lambda_3(R) <0$ for all $r>0$ and $g>0$. Since $\lambda_1(R) = -1$
this demonstrates that the racemic state $R$ is always stable. As an
independent check, we also verify that $\lambda_3(Q) > 0$ is
positive for all $r>0$ and $g>0$, so the chiral solutions $Q_{\pm}$
are always unstable. There is no stable mirror symmetry broken
solution when the heterodimers (formed through the crucial mutual
inhibition step) are \textit{included} reversibly in the system, see
Table \ref{tab:table1}.

\subsection{\label{sec:Frankrevhomhet} Frank with reversible homo- and heterodimer formation}

If heterodimers can form, it is certainly reasonable to expect the
same for homodimers. This corresponds to the reaction steps
Eqs.(\ref{decay},\ref{autoLD2},\ref{homodim},\ref{heterodim}), and
we must solve all five equations
Eqs.(\ref{chi},\ref{y},\ref{chi2},\ref{y2},\ref{P}) after setting
$w=0$ and $s=1$. To help maintain algebraic control, we also will
set $u=0$. There are four static solutions:
\begin{eqnarray}\begin{array}{lll}
O & \equiv (P = 0, \chi_2=0, y_2=0, \chi=0, y = 0) \\
R & \equiv (P= \frac{1}{(g-q)^2 r},\chi_2 = \frac{q}{p(g-q)^2}, y_2 = 0, \chi = \frac{2}{g-q}, y = 0) \\
Q_{\pm}& \equiv \left(P=0, \chi_2 = \frac{q}{2p(g-q)^2},y_2 =
\pm\frac{q}{2p(g-q)^2}, \chi = \frac{1}{g-q}, y= \pm\frac{1}{g-q}
\right).
\end{array}
\end{eqnarray}
It is important to point out that $g=\frac{k_{-2} + 2k_4}{k_5} =
\frac{k_{-2}}{k_5} + q$, so that $g > q$.  Note that the final
heterodimer concentration $P$ is \textit{zero} in the chiral states
$Q_{\pm}$, but there a net positive concentration $\chi_2 > 0$ for
the homodimers.

In order to study the stability of the four possible homogeneous
solutions $O, R$ and $Q_{\pm}$, we proceed to calculate the
eigenvalues of the full $5\times 5$ array in Eq.(\ref{Mopen}). As it
turns out, we are unable to obtain analytic closed-form expressions
for the eigenvalues of the racemic state $R$, but we can do so for
both the empty $O$ and the chiral solutions $Q_{\pm}$, which is
sufficient for our purposes. These are given by:
\begin{eqnarray}
  \lambda_{1,2,3,4,5}(O) &=& (1,1,-p,-p,-r) \\
  \lambda_{1,2,3,4,5}(Q_{\pm}) &=& \Big( -p,-\frac{g+g p+q-p q+\sqrt{-4 p (g-q)^2+(g+g p+q-p q)^2}}{2 (g-q)},
  \nonumber \\
&& \frac{-g (1+p)+(-1+p) q+\sqrt{-4 p (g-q)^2+(g+g p+q-p q)^2}}{2
(g-q)}, \nonumber \\
&&-\frac{1+q+g (-1+r)-q r+\sqrt{4 (g-q)^2 r+(1+q+g (-1+r)-q r)^2}}{2 (g-q)}, \nonumber \\
&& \frac{-1+g-q-g r+q r+ \sqrt{4 (g-q)^2 r+(1+q+g (-1+r)-q r)^2}}{2
(g-q)}\Big).
\end{eqnarray}
As before, the empty solution is always unstable. As for the chiral
solution, an inequality analysis shows that $\lambda_{1,2,3,4}(Q) <
0$ are all negative, whereas $\lambda_5(Q)
>0$. Hence the chiral asymmetric solutions are unstable, and thus by logical
deduction, the racemic state $R$ must therefore always be stable. We
point out however, that if the heterodimers are formed
\textit{irreversibly} $(r=0)$, then a stable homochiral outcome is
again possible whenever $g < 1$.

To summarize up to this point, the sequence of Frank type models
analyzed above indicates that the elimination of the heterodimers
from the system (when $[A]$ is constant), or their irreversible
formation, is crucial in order that symmetry breaking be possible
and that the chirally asymmetric states be stable. Allowing for
homodimer formation does not alter the stability properties. See
also the first four rows of Table \ref{tab:table1}.

\subsection{\label{sec:limenant}Limited enantioselectivity}

Whereas the forward reactions in Eq.(\ref{autoLD2}) represent the
autocatalytic capabilities of each enantiomer, the forward reaction
in Eq.(\ref{nonenantiosel}) accounts for the fact that an enantiomer
can also catalyze the production of its chiral partner. This
scenario, corresponding to the reaction steps
Eqs.(\ref{decay},\ref{autoLD2},\ref{nonenantiosel}), is termed
\textit{limited enantioselectivity }\cite{Avet,GH}. As there is no
dimerization in this model, we are to solve Eqs.(\ref{chi},\ref{y})
in which we set $\chi_2=P=y_2=0$. In this case, the four static
solutions are (note we also set $u=0$ to help simplify the algebra):
\begin{eqnarray}\begin{array}{lll}
O & \equiv (\chi =0 , y = 0) \\
R & \equiv (\chi = {2(1 + w) \over 1 + g} , y = 0) \\
Q_{\pm}& \equiv \left(\chi = {1-w \over g}  , y = \pm
\sqrt{{(1-w)(g+3wg+w -1) \over (g-1)g^2}}\, \right).
\end{array}
\end{eqnarray}
$O$ denotes the empty solution, that is, with zero chiral matter,
$R$ is the racemic solution with positive net total chiral matter,
and $Q_{\pm}$ denote the two possible chiral solutions. Because
there is no dimerization in this model, we can set $k_4 = k_5=0$ in
Eq.(\ref{openparams1}) and so we find $g = \frac{k_{-2}}{k_{-3}}$.

\begin{figure}[ht]
 \includegraphics[width=10.0cm]{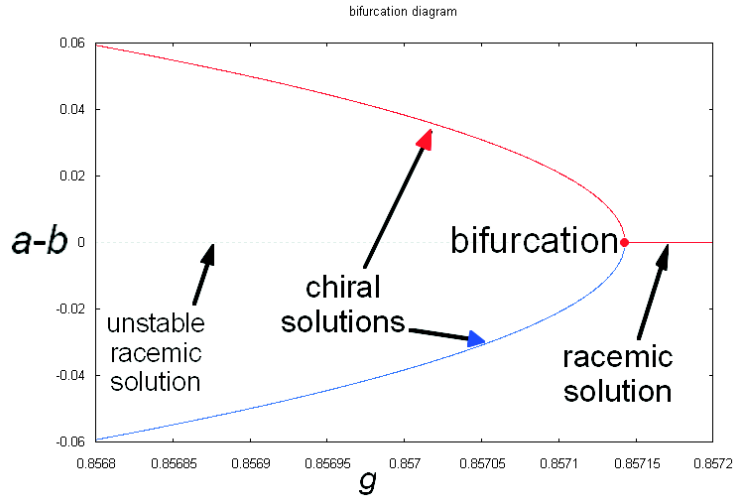}
  \caption{\label{bifurc}Bifurcation diagram, limited enantioselectivity. For
  $g>\frac{1-w}{1+3w}$,
  the unique solution is $y=0$, i.e. $[L]=[D]$, which corresponds to a racemic solution.
  Reducing $g$ below the critical value, the system undergoes
  a pitchfork bifurcation, leading to chiral solutions, i.e. $[L] \neq
  [D]$. This is illustrated here for $g_{crit} = 0.85714$ ($w =
  0.04$).
  These chiral solutions however are not homochiral, as they yield an enantiomeric excess $ee$ strictly less than
  unity, provided that $k_3 \neq 0$; see Eq.(\ref{eeopen}).}\label{bif}
\end{figure}

In order to study the stability of the four possible homogeneous
solutions $O, R$ and $Q_{\pm}$, we calculate the eigenvalues of the
$2\times 2$ subblock of the upper left hand corner of the Jacobian
matrix Eq.(\ref{Mopen}) evaluated at each one of these four possible
solutions. The eigenvalues are given by
\begin{eqnarray}
  \lambda_{1,2}(O) &=& (1-w,1+w) \\
  \lambda_{1,2}(R) &=& (-1-w,\frac{1-w-g-3 w g}{1+g})\\
  \lambda_{1,2}(Q_{\pm}) &=& \left(\frac{-1+w+2w g - \sqrt{(1-w)^2+g^2(4+8w-8w^2)-4g(1-w)}}{2
  g} \right.  \nonumber \\
  &&\left.\frac{-1+w+2w g + \sqrt{(1-w)^2+g^2(4+8w-8w^2)-4g(1-w)}}{2
  g}\right).\nonumber \\
  & &
\end{eqnarray}
The empty solution $O$ is always unstable. To ensure the stability
in $R$, we need for both eigenvalues to be negative:
$\lambda_{1,2}(R) <0$. While the first eigenvalue is always
negative, we have $\lambda_{2}(R) <0$ if and only if
$g>\frac{1-w}{1+3w}$. Concerning the stability of the solutions
$Q_{\pm}$, these are always unstable if $g>1$ and they become stable
for $0<g<\frac{1-w}{1+3w}<1$. When this holds, the final
enantiomeric excess $ee$ will be given by
\begin{equation}\label{eeopen}
ee = \pm\sqrt{1-\frac{4gw}{(1-w)(1-g)}},
\end{equation}
and so $0< |ee| < 1$, which means that the chiral solution is not
homochiral. This symmetry breaking can be represented via a
corresponding bifurcation diagram, see Fig \ref{bifurc}. When
$g>g_{crit}$ only the racemic state is stable, and $ee=0$, but when
$g<g_{crit}$ two stable chirally asymmetric states are equally
likely, which one the system actually chooses is a random event. The
upper and lower branches of the bifurcation are given by plotting
the values of $ee$, Eq.(\ref{eeopen}) as a function of $g$ holding
$w$ fixed.

In summary, there is a critical value of the parameter,
\begin{equation}\label{gcritopen}
g_{crit} = \frac{1-w}{1+3w},
\end{equation}
that uniquely determines the outcome of the reaction scheme. For
$g>\frac{1-w}{1+3w}$ the only stable solution is $R$. In this case
$y=0$ i.e. $[L]=[D]$, i.e. we have a \textit{racemic} solution. Note
that if the decay $k_{-1}$ of the enantiomers into achiral matter
can be neglected with respect to the rate of autocatalytic
amplification, then $w$ is well approximated by the ratio $k_3/k_2$,
independent of the concentration of achiral matter $[A]$, and so the
critical value $g_{crit}$ and final solution are controlled by the
competition between the rates of the forward reactions Eqs.(2.2) and
(2.3). That is, the relative rate of autocatalysis versus limited
enantioselectivity, in which the catalytic effect of each enantiomer
leads to the formation of \textit{both} L and D products. On
thermodynamic grounds however, the symmetry breaking condition $g <
g_{crit}$, determined from the stability analysis, cannot be
achieved \cite{GH}. Enantiomers are thermodynamically identical
species and therefore, at the final state must fulfill the condition
$\frac{k_1}{k_{-1}} = \frac{k_2}{k_{-2}} = \frac{k_3}{k_{-3}}$,
which is a consequence of the principle of detailed balance
\cite{Lebon}. Therefore, the necessary condition $g<1$ (i.e.,
$k_{-3} > k_{-2})$ is incompatible with $k_2 > k_3$, and $g$ will
always be greater than the critical value. The question then arises
if it is possible or not to find more complex non-linear reaction
networks where the mathematical condition for symmetry breaking, as
determined by the stability analysis, is compatible with
thermodynamic chemical constraints.

\begin{table}
\caption{\label{tab:table1} Selected reaction schemes treated as
open systems, and the corresponding condition $g<g_{crit}$ that must
be satisfied for obtaining a stable chiral solution.}
\begin{ruledtabular}
\begin{tabular}{cccc}
Reaction model (\textbf{open systems}: $[A]=$ constant) & $g$  & $g_{crit}$  &  $g<g_{crit}$   \\
\hline
Frank (with $k_1=k_{-1}= 0$)\footnotemark[1]&  $\frac{k_{-2}}{k_5}$ & 1   &  $k_{-2} < k_5$ \\
  &  &  & \\
  Frank (with $k_1>0,\,k_{-1}> 0$)&  $\frac{k_{-2}}{k_5}$ & $\frac{1}{8u}(\sqrt{1+16u}-1)$\footnotemark[2]
  & $\frac{k_{-2}}{k_5} < \frac{1}{8u}(\sqrt{1+16u}-1)$ \\
  &  &  & \\
Frank + \textit{reversible} heterodimers \footnotemark[3]& $\frac{k_{-2}}{k_5}$ & 0\,\footnotemark[4] & -- \\
  &  &  & \\
  Frank + \textit{reversible} homo- and heterodimers \footnotemark[5]& $\frac{k_{-2}+2k_4}{k_5}$ & $q=\frac{2k_4}{k_5}$ \, &
   -- \footnotemark[6]\\
  &  &  & \\
Limited enantioselectivity \footnotemark[7]& $\frac{k_{-2}}{k_{-3}}$
& $\frac{1-w}{1+3w}$\,\footnotemark[8]
& $\frac{k_{-2}}{k_{-3}} < \frac{1-w}{1+3w}$ \\
  &  &  & \\
Lim. enantiosel. plus heterodimers & $\frac{k_{-2}}{k_{-3}+k_5}$ &
$(1-s)\frac{1-w}{1+3w}$\,\footnotemark[9] &  $\frac{k_{-2}}{k_{-3}} < \frac{1-w}{1+3w}$ \\
  &  &  & \\
Lim. enantiosel. plus homodimers & $\frac{k_{-2}+ 2k_4}{k_{-3}}$ &
$\frac{1+q-w+3qw}{1+3w}$\,\footnotemark[10] &  $\frac{k_{-2}}{k_{-3}} < \frac{1-w}{1+3w}$\\
  &  &  & \\
Lim. enantiosel. plus homo- and heterodimers & $\frac{k_{-2}+
2k_4}{k_{-3}+k_5}$ &
$\frac{1+q-s-w+3qw+sw}{1+3w}$\,\footnotemark[11]  & $\frac{k_{-2}}{k_{-3}} < \frac{1-w}{1+3w}$ \\
\end{tabular}
\end{ruledtabular}
\footnotetext[1]{Reaction steps Eqs.(\ref{autoLD2},\ref{heterodim});
see Sec \ref{sec:Frank}. The system is fed by an input of the
achiral substrate $A$, and the output consists of the $\rm
LD$-heterodimers formed in the mutual inhibition step
Eq.(\ref{heterodim}); see e.g., Ref. \cite{Frank, Mason,GTV}.}
\footnotetext[2]{Reaction steps
Eqs.(\ref{decay},\ref{autoLD2},\ref{heterodim}); see Sec
\ref{sec:Frank}. Here, $u = \frac{k_1[A]k_5}{(k_2[A]-k_{-1})^2}$}.
\footnotetext[3]{Reaction steps Eqs.(\ref{autoLD2},\ref{heterodim});
see Sec \ref{sec:Frankrevhet}. The heterodimer concentration is
\textit{time dependent}, in contrast to the original Frank model.}
\footnotetext[4] {$g_{crit} = 1-s$, but in this model $k_3 =
k_{-3}=0$, so that $s = k_5/k_5 =1$.} \footnotetext[5]{Reaction
steps Eqs.(\ref{autoLD2},\ref{homodim},\ref{heterodim}); see Sec
\ref{sec:Frankrevhomhet}. Both homo and heterodimers are formed, and
are \textit{not} removed from the system. } \footnotetext[6]{In this
model, $g \geq q = 2k_4/k_5$, so no symmetry breaking is possible. }
\footnotetext[7]{Reaction steps
Eqs.(\ref{autoLD2},\ref{nonenantiosel}), see Sec
\ref{sec:limenant}.} \footnotetext[8]{Recall $w = k_3/k_2$ for large
constant concentrations $[A]$.} \footnotetext[9]{Reaction steps
Eqs.(\ref{autoLD2},\ref{nonenantiosel},\ref{heterodim}), see Sec
\ref{sec:limenantdim}. In this model $s=k_5/(k_{-3}+k_5) < 1$.}
\footnotetext[10]{Reaction steps
Eqs.(\ref{autoLD2},\ref{nonenantiosel},\ref{homodim}), see Sec
\ref{sec:limenantdim}. In this model, since $k_5=0$, then
$q=2k_4/k_{-3}$.} \footnotetext[11]{Reaction steps
Eqs.(\ref{autoLD2}-\ref{heterodim}), see Sec \ref{sec:limenantdim}.
Here $q=2k_4/(k_{-3}+k_5)$ and $s=k_5/(k_{-3}+k_5) < 1$ for $k_{-3}
> 0$.}
\end{table}
%

\subsection{\label{sec:limenantdim} Limited enantioselectivity plus dimers}

It is natural to ask if including dimer formation can alter the
symmetry breaking condition $g < g_{crit}$ and get around the
thermodynamic constraint. To this end, and for more chemical
realism, we include the dimerizations. We begin with just the
hetero-dimerization step, and the static solutions work out to be
given by
\begin{eqnarray}
O &\equiv& \big(P = 0, y = 0, \chi = 0 \big) \\
R &\equiv& \big( P = \frac{s (1+w)^2}{r (1+g-s)^2}, y = 0, \chi =
\frac{2 (1+w)}{1+g-s} \big) \\
Q_{\pm} &\equiv& \big( P = \frac{s (-1+w) w}{g r (-1+g+s)}, y = \pm
\frac{\sqrt{1-w} \sqrt{\frac{-1+g+s+w+3 g w-s w}{g}}}{\sqrt{g}
\sqrt{-1+g+s}}, \chi = \frac{1-w}{g} \big)
\end{eqnarray}

The eigenvalues for the empty and racemic solution can be found in
closed form. They are given by
\begin{equation}
\lambda_{1,2,3}(O) = (-r,1-w,1+w).
\end{equation}
and
\begin{eqnarray}
&& \lambda_{1,2,3}(R) = \Big( 1-w-\frac{2 g (1+w)}{1+g-s}, \nonumber
\\
&-& \frac{1+r+s-r s+w+s w+g (1+r+w)}{2 (1+g-s)} \nonumber \\
&+& \frac{\sqrt{-4 r (1+g-s)^2 (1+w)+(1+r+s-r s+w+s w+g
(1+r+w))^2}}{2
(1+g-s)},\nonumber \\
& & \frac{r (-1-g+s)-(1+g+s) (1+w)}{2 (1+g-s)} \nonumber
\\
&+& \frac{\sqrt{-4 r (1+g-s)^2 (1+w)+(1+r+s-r s+w+s w+g
(1+r+w))^2}}{2 (1+g-s)} \Big).
\end{eqnarray}
Note that here, $g =\frac{k_{-2}}{k_{-3}+k_5}$. The empty solution
$O$ is always unstable. As for the racemic solution $R$, we note
that $\lambda_1(R) < 0$ provided that $g > \frac{(1-s)(1-w)}{1+3w}$.
The remaining two eigenvalues $\lambda_{2,3}(R) < 0$ for
$r>0,w>0,1>s>0$ and for all $g>0$. So the racemic state is stable
when $g > g_{crit} = \frac{(1-s)(1-w)}{1+3w}$ and unstable when $g <
g_{crit}$. Clearly, when the racemic state is unstable, we infer
that chiral state is the only stable solution.

Note that the expression for the critical $g$ in this model is just
the $g_{crit}$ for the limited enantioselectivity model treated
above times the additional factor $(1-s)$, see Table
\ref{tab:table1}. In other words, the inclusion of heterodimer
formation and dissociation seems to \textit{reduce} the value of
$g_{crit}$, appearing to making chiral symmetry breaking relatively
more difficult to achieve. But since $s=\frac{k_5}{k_{-3}+k_5}$, the
crucial inequality $g < g_{crit}$ actually reduces algebraically and
simplifies to yield the one obtained for the limited
enantioselectivity model without heterodimers.

For completeness, we also have worked out the fixed point solutions
and their stability properties first, when only the homodimers, and
then when both the homo-and heterodimers are included. The results
of these latter two cases are summarized in the seventh and eighth
rows of Table \ref{tab:table1}. It is interesting to point out that
although the expressions for the critical $g$'s depend on the homo
and heterodimer rates explicitly and in a non-trivial way (see the
third column), the final inequalities $g < g_{crit}$ determining the
stable chiral outcomes are the same as for limited
enantioselectivity alone, irrespective of whether the homo and/or
heterodimers are included. Compare the fifth and seventh rows of the
Table, as well as the sixth and eighth rows. The dependence on $k_4$
or $q$ simply cancels out. The inclusion of dimer formation does not
affect the symmetry breaking in this model. Since the condition
$\frac{k_1}{k_{-1}} = \frac{k_2}{k_{-2}} = \frac{k_3}{k_{-3}}$
continues to hold, only the racemic state is stable.
%

\section{\label{sec:closedsystems}Closed reaction systems}

For closed systems there is no flow of material into or out of the
system. Since $[A]$ is not constant in this situation, we cannot use
it to rescale the time or the concentrations. We can however take
$\tau = k_1 t$ for the time and $[\tilde L] = \frac{k_{-3} +
k_5}{k_1} [L]$, etc. for the dimensionless concentrations. This
allows us to express the rate equations in Sec \ref{sec:mainmodel}
in the following dimensionless form:
\begin{eqnarray}\label{eqntilL}
\frac{d}{d \tau}[\tilde L] &=& [\tilde A] - u[\tilde L] + h[\tilde
A][\tilde L] - g[\tilde L]^2 -[\tilde L][\tilde D]
+ r[\tilde A][\tilde D]  + 2\omega[\tilde L_2] + \rho[\tilde{LD}], \nonumber \\
\label{eqntilD} \frac{d}{d \tau}[\tilde D] &=& [\tilde A] - u[\tilde
D] + h[\tilde A][\tilde D] - g[\tilde D]^2 -[\tilde D][\tilde L] +
r[\tilde A][\tilde L]  + 2\omega[\tilde D_2] + \rho[\tilde{LD}],
\nonumber \\
\frac{d}{d \tau}[\tilde{L}_2] &=& \frac{q}{2}[\tilde L]^2 -
\omega[\tilde{L}_2], \nonumber \\
\frac{d}{d \tau}[\tilde{D}_2] &=& \frac{q}{2}[\tilde D]^2 -
\omega[\tilde{D}_2], \nonumber \\
\frac{d}{d \tau}[\tilde{LD}] &=& s[\tilde L][\tilde D] -
\rho[\tilde{LD}]. \nonumber
\end{eqnarray}
These are subject to the constraint $[\tilde A] = C - [\tilde
L]-[\tilde D]- 2[\tilde{L}_2] -2[\tilde{D}_2] -2[\tilde{LD}]$, where
$C$ is a constant. The parameters appearing here are
\begin{eqnarray}\label{paramsclosed}
u &=& \frac{k_{-1}}{k_1},\, g =\frac{k_{-2}+ 2k_4}{k_{-3}+k_5}, \, h
= \frac{k_2}{k_{-3}+k_5}, \, r = \frac{k_3}{k_{-3}+k_5}, \, \rho =
\frac{k_{-5}}{k_1}, \, \omega=\frac{k_{-4}}{k_1}, \,
q=\frac{2k_4}{k_{-3}+k_5}, \nonumber \\
&{\rm and}& \, s= \frac{k_{5}}{(k_{-3}+ k_5)}.
\end{eqnarray}
For closed systems, the number of independent parameters reduces
from ten to eight.

As before, we find it convenient to define the sums and differences
of concentrations: $\chi = [\tilde L] + [\tilde D]$, $y = [\tilde L]
- [\tilde D]$, $\chi_2 = [\tilde L_2] + [\tilde D_2]$ $y_2 = [\tilde
L_2] - [\tilde D_2]$, and put $P = [\tilde{LD}]$. This then yields
the following:

\begin{eqnarray}\label{closedchi}
\frac{d}{d \tau}\chi &=& 2[\tilde A] + ((h+r)[\tilde A]-u)\chi
-\frac{1}{2}(g+1)\chi^2 -\frac{1}{2}(g-1)y^2 + 2\omega \,\chi_2 +
2\rho \,P,\\\label{closedy} \frac{d}{d \tau}y &=& y
\big((h-r)[\tilde A] -g\chi-u \big) + 2\omega \,y_2,
\\\label{closedchi2} \frac{d}{d \tau} \chi_2 &=&  \frac{q}{4}(\chi^2
+ y^2) - \omega \, \chi_2,
\\\label{closedy2}
\frac{d}{d \tau} y_2 &=&  \frac{q}{2}\chi y - \omega \, y_2,
\\
\label{closedP} \frac{d}{d \tau} P &=&  \frac{s}{4}(\chi^2-y^2) -
\rho P.
\end{eqnarray}
In these variables, the constant mass constraint reads $[\tilde A] =
C - \chi -2 \chi_2 - 2P$.

For the complete closed model, that is, the reaction steps in
Eqs.(\ref{decay}-\ref{heterodim}), the $5 \times 5$ Jacobian matrix
$M_{closed}$ for the linearized fluctuation equations is given by
\begin{equation}\label{Mclosed}
M_{closed} = \left(
  \begin{array}{ccccc}
    \scriptstyle{(h+r)(C-2\chi_2 - 2P)-(2+u)} & \scriptstyle{(1-g)y} & \scriptstyle{2\omega-4-2(h+r)\chi} & 0 &
    \scriptstyle {2\rho-4-2(h+r)\chi} \\
      \scriptstyle{-(g+1+2(h+r))\chi}  &    &     &    &     \\
        &    &     &    &     \\
    \scriptstyle{-(g+h-r)y} & \scriptstyle{(h-r)(C-2\chi_2-2P)} & \scriptstyle{-2(h-r)y} & \scriptstyle{2\omega} &
    \scriptstyle{-2(h-r)y} \\
        & \scriptstyle{-(g+h-r)\chi -u}     &    &      &     \\
        &     &    &      &     \\
    \frac{q}{2}\chi & \frac{q}{2}y&-\omega & 0 & 0 \\
    \frac{q}{2}y & \frac{q}{2}\chi & 0 & -\omega & 0 \\
    \frac{s}{2}\chi & -\frac{s}{2}y & 0 & 0 &-\rho \\
  \end{array}
\right).
\end{equation}
This incorporates the constraint equation directly. This matrix is
to be evaluated on any of the static fixed point solutions
$\chi^*,y^*,\chi^*_2,y^*_2$ and $P^*$ of the equation set
Eqs.(\ref{closedchi}-\ref{closedP}). The constant $C$ is related to
the total initial concentration $Q$ as follows: $C =
\frac{k_{-3}+k_5}{k_1} Q$. The eigenvalues of $M_{closed}$ indicate
the stability of the static solution on which this matrix is
evaluated. Unlike the open flow case however, it is much more
difficult to obtain manageable analytic expressions for these
eigenvalues, and the variety of reaction schemes we can treat
analytically in this manner is limited.

\subsection{Closed Frank with irreversible heterodimers: $k_{-5}= 0$}
Recall that the Frank model in an open flow system and with
irreversible formation of heterodimers does lead to stable
steady-state chiral solutions. We now enclose it in a box, so that
$[A]$ is no longer constant. In this situation we are to solve the
Eqs.(\ref{closedchi},\ref{closedy},\ref{closedP}) where from the
parameter list in Eq(\ref{paramsclosed}) we have $u>0$,
$g=k_{-2}/k_5$, $h=k_2/k_5$, $r=\omega=q=0$, $s=1$ and for
$k_{-5}=0$, we must set $\rho = 0$. The mass constraint reads
$[\tilde A] = C-\chi-2P$. There is a static racemic $R$ and two
static chiral $Q_{\pm}$ solutions:
\begin{eqnarray}\begin{array}{lll}
R & \equiv (P= \frac{C}{2}, y = 0, \chi= 0) \\
Q_{\pm}& \equiv \left(P = \frac{Cg+u}{2g} , y = \pm \frac{u}{g},
\chi = -\frac{u}{g}
 \right).
\end{array}
\end{eqnarray}
Note that the chiral solutions are unphysical since they imply $\chi
< 0$ when $u>0$.  We can attempt to get some information regarding
stability by evaluating the $3\times 3$ array obtained from
Eq.(\ref{Mclosed}) after deleting the 3rd and 4rth rows and columns.
The eigenvalues corresponding to the racemic solution are
\begin{equation}
\lambda_{1,2,3}(R) = (0,-2-u,-u),
\end{equation}
indicating that $R$ is marginally stable (\textit{marginal} because
of the zero entry). We are unable to calculate the eigenvalues for
the unphysical chiral states in simple analytic form. For $u=0$, the
only steady state is the racemic one $R$, (there is no steady chiral
solution) and its eigenvalue is (0,-2,0).

This is important, because Model 1 of Rivera Islas et al
\cite{Islas}, is actually a special case of this, in which $u =
g=0$, and we consider this next.

\subsection{Rivera Islas  et. al. Model 1}

This model \cite{Islas} corresponds to our reaction steps
Eqs(\ref{decay},\ref{autoLD2},\ref{heterodim}) and after setting
$k_{-1} = k_{-2} = k_{-5} = 0$. Note this model implies
\textit{irreversible} reactions. This is a closed system, we we
solve again the Eqs.(\ref{closedchi},\ref{closedy},\ref{closedP})
where from the parameter list in Eq(\ref{paramsclosed}), we now have
$u = 0$, $g=0$, $h=k_2/k_5$, $r=\omega=q=0$, $s=1$ and for
$k_{-5}=0$, we must set $\rho = 0$. The constraint is $[\tilde A] =
C-\chi-2P$. We next look for the static solutions of the set of
equations Eqs.(\ref{closedchi},\ref{closedy},\ref{closedP}). The
racemic solution $y=0$ implies that $\chi = 0$ and $P = C/2$. Thus,
all the net matter ends up as heterodimers. On the other hand, the
chiral solution has $y \neq 0$. For this case, the static solution
of Eqs.(\ref{closedchi},\ref{closedy},\ref{closedP}) yields $\chi^2
= y^2$ and $C-2P = \chi$. The physical solution corresponds to a
positive $\chi >0$, so that $y = \pm\chi $ and $P=\frac{C-\chi}{2}
>0$. In this case, the total matter is distributed among monomers
and dimers. Note however we cannot independently solve for the final
heterodimer and monomer concentration in the chiral phase.  The
static solutions are summarized as follows:
\begin{eqnarray}\begin{array}{lll}
R & \equiv (P= \frac{C}{2}, y = 0, \chi= 0) \\
Q_{\pm}& \equiv \left(P = \frac{C-\chi}{2} , y = \pm \chi \right).
\end{array}
\end{eqnarray}

Using $M_{closed}$ to compute the eigenvalues for this case, we find
that
\begin{equation}
{\lambda_{1,2,3}(R)}_{Islas} =(-2,0,0).
\end{equation}
Since ${\lambda_{2,3}(R)}_{Islas} =0$, we cannot say conclusively if
this state is stable or unstable. This is in accord with the
chemical fact that due to its irreversibility, this reaction scheme
leads to a kinetic controlled outcome of the reaction products.
Evaluating Eq.(\ref{Mclosed}) on this chiral solution and
calculating the eigenvalues yields
\begin{equation}
{\lambda_{1,2,3}(Q)}_{Islas} = ( 0,-\chi, -2-h\chi).
\end{equation}
This is as much as we can say regarding linear stability analysis;
there are \textit{no free} adjustable parameters that can induce a
change in sign in any of the eigenvalues. The simple stability
analysis is inconclusive. Nevertheless, numerical simulations
carried out in \cite{Islas} indicate stable chirally asymmetric
solutions for a range of $h$. Thus, a higher order stability
analysis might be called for.

\subsection{Limited enantioselectivity revisited: the closed system}

We can obtain the static solutions and their associated eigenvalues
exactly for limited enantioselectivity in a closed system, the model
introduced in Subsection \ref{sec:limenant}. This is useful because
it provides an exact point of comparison between a specific set of
reactions in both open and closed systems. So we consider the
reaction set Eqs.(\ref{decay},\ref{autoLD2},\ref{nonenantiosel}),
but now in a closed system. We must solve
Eqs.(\ref{closedchi},\ref{closedy}) in which we set
$\chi_2=P=y_2=0$. The mass constraint in this case reads $[\tilde A]
= C - \chi$. There are four static solutions (note we also set $u=0$
to simplify somewhat the algebra):
\begin{eqnarray}
U \equiv \big(y &=& 0, \, \chi = -\frac{2-C (h+r)+\sqrt{4 C (1+g+2(
h+r))+(2-C
(h+r))^2}}{1+g+2(h+r)} \big) \ \\
R \equiv \big(y &=& 0, \, \chi = \frac{-2+C (h+r)+\sqrt{4 C (1+g+2
(h+r))+(2-C (h+r))^2}}{1+g+2(h+r)}\big) \ \\\label{Q} Q_{\pm} \equiv
\big( y &=& \pm \sqrt{\frac{4 C g (g+h-r)+C^2 (h-r) ((-1+g) h+r+3 g
r)}{(g-1)(g+h-r)^2}}, \,\chi = \frac{C(h-r)}{g+h-r}\big).
\end{eqnarray}
$U$ denotes the \textit{unphysical} solution, as this implies a
negative total chiral matter, $R$ is the racemic solution with
positive net total chiral matter, and $Q_{\pm}$ denote the two
possible chiral solutions. Note from Eq.(\ref{paramsclosed}) we have
$g =\frac{k_{-2}}{k_{-3}}, \, h = \frac{k_2}{k_{-3}}, \, r =
\frac{k_3}{k_{-3}}$.

The stability of the four possible homogeneous solutions $U, R$ and
$Q_{\pm}$, is determined from considering the eigenvalues of the
$2\times 2$ subblock of the upper left hand corner of the Jacobian
matrix $M_{closed}$ Eq.(\ref{Mclosed}) (and after setting $\chi_2 =
P=0$ as well as $u=0$) evaluated at each one of these four possible
solutions. The eigenvalues are given by

\begin{eqnarray}
\lambda_{1,2}(U) &=& \Big(C (h-r)-\frac{(g+h-r)}{1+g+2 h+2
r}\left(-2+C (h+r)-\sqrt{4+C^2 (h+r)^2+4 C (1+g+h+r)}\right),
\nonumber \\
& & \sqrt{4+C^2 (h+r)^2+4 C (1+g+h+r)} \Big)
\end{eqnarray}
\begin{eqnarray}
\lambda_{1,2}(R) &=& \Big(C (h-r)-\frac{(g+h-r)}{1+g+2 h+2 r}
\left(-2+C (h+r)+\sqrt{4+C^2 (h+r)^2+4 C (1+g+h+r)}\right),
\nonumber \\
&-& \sqrt{4+C^2 (h+r)^2+4 C (1+g+h+r)} \Big)
\end{eqnarray}

and
\begin{eqnarray}
\lambda_{1,2}(Q_{\pm}) &=& \frac{1}{2 (g+h-r)}(2 g (-1+C r)+(-h+r)
(2+C (1+h+r))\nonumber \\
&+& \surd ((g (2-2 C r)+(h-r) (2+C (1+h+r)))^2 \nonumber \\
&+& 4 C (g+h-r) (4 g^2-C (h-r)^2+g (h-r) (4+C (h+3
r))))),\nonumber \\
&& \frac{1}{2 (g+h-r)}(2 g (-1+C r)+(-h+r)
(2+C (1+h+r))\nonumber \\
&-& \surd ((g (2-2 C r)+(h-r) (2+C (1+h+r)))^2 \nonumber \\
&+& 4 C (g+h-r) (4 g^2-C (h-r)^2+g (h-r) (4+C (h+3 r)))))\nonumber
\end{eqnarray}
As $\lambda_2(U) > 0$, the unphysical solution $U$ is therefore
always unstable. To ensure the stability in $R$, we need for both
eigenvalues to be negative: $\lambda_{1,2}(R) <0$. While the second
eigenvalue is always negative, we have $\lambda_{1}(R) <0$ if and
only if $g > g_{crit}$, where
\begin{equation}\label{gcritclosed}
g_{crit} = -\frac{1}{8}(h-r) \left(4+C (h+3 r)-\sqrt{16+C^2 (h+3
r)^2+8 C (2+h+3 r)}\right).
\end{equation}
A careful expansion of this shows that in the limit of  large $C$,
then the critical value of $g$
\begin{equation}
g_{crit} \rightarrow \frac{1-r/h}{1+3r/h} =
\frac{1-k_3/k_2}{1+3k_3/k_2} = \frac{1-w}{1+3w} = g_{crit}^{open},
\end{equation}
asymptotically approaches the critical $g$ for the same set of
reactions in an open system, Eq.(\ref{gcritopen}).

Concerning the stability of the solutions $Q_{\pm}$, these are
always unstable if $g>1$ and they become stable
$\lambda_{1,2}(Q_{\pm}) < 0$ for $0<g<g_{crit}<1$. However, the
thermodynamic condition mentioned above still holds and implies that
the racemic state is the only stable outcome.
%

\section{\label{sec:disc} Concluding remarks}

We have focussed attention on the steady state solutions and their
dynamic stability properties in variants of both the Frank and
limited enantioselectivity models of mirror symmetry breaking. In
the Frank model \cite{Frank,Mason}, the mutual inhibition occurs
through the formation of heterodimers composed of the two chiral
monomers, whereas in limited enantioselectivity \cite{Avet,GH}, the
needed chiral antagonism occurs though a monomer recycling reaction.
We have unified both models into one encompassing reaction scheme
and have studied the steady state solutions and their stability
properties in a sequence of reaction schemes that start from the
paradigmatic Frank model to limited enantioselectivity, in which we
study the consequences of including the formation of (reversible)
hetero- and homo-dimers as dynamic variable concentrations in both.
The general conclusion we can draw from this is that the inclusion
of the variable heterodimer concentrations leads to a final stable
racemic state in the sequence of Frank-type models defined above.
This is important because we recall that in Frank's original
formulation, the heterodimers are supposed to be \textit{removed}
from the system, i.e., their formation is irreversible. Including
the homodimers does not affect this conclusion.

In the case of the limited enantioselectivity model, no dimer
formation is originally contemplated, and mirror symmetry breaking
is \textit{mathematically} possible because the needed inhibition is
provided by the reverse symmetric catalysis step \cite{GH}. In spite
of this, the symmetry breaking condition $g < g_{crit}$, determined
from the pure stability analysis, cannot be achieved on
thermodynamic grounds \cite{GH}, as has also been pointed out
recently \cite{Matar}. Including the homo- and/or the heterodimer
formation to limited enantioselectivity does not alter the crucial
inequality, nor does the presence of the dimers change the
thermodynamic constraint. Therefore an open question is if a
different chemical scenario leading to collective phenomena exists,
such as a second order phase transition, that can give rise to the
condition $g < g_{crit}$. In the context of spatially extended
polymerization systems, it has been argued before that one should
expect domain formation similar to the case of second order phase
transitions \cite{BM,Gleiser,GW}.

The reaction networks treated here are limited to reactions
involving only monomers and dimers, whereas biological chirality of
living systems involves large macromolecules that are probably the
result of polimerization reactions. In this vein, Sandars recently
introduced a model in which the detailed polymerization process and
enantiomeric cross-inhibition are taken into account, its basic
features are explored numerically, but without including spatial
extent, chiral bias or noise \cite{Sandars}. Brandenburg and
coworkers have analyzed further properties of Sandars' model and
have proposed a truncated version including chiral bias \cite{BAHN},
and have studied this reduction with spatial extent and coupling to
a turbulent advection velocity \cite{BM}. Gleiser and Thorarinson
analyze the reduced Sandars' model with spatial extent and coupling
to an external white noise \cite{GT} and in \cite{Gleiser}, Gleiser
considers the reduced chiral biased model with external noise. In
addition to Sandars, both Wattis and Coveney \cite{Wattis} and Saito
and Hyuga \cite{Saito} have introduced polymerization models that
can give rise to homochiral states. The latter one differs from
Sandars' in allowing for reversibility in all the steps.  As
reported in the review article (d) in \cite{all}, all these
polymerization models derive from Frank's model by adding
polymerization reactions. Thus, in spite of the simplicity of the
original Frank model \cite{Frank}, ignoring as it does the
polymerization process, it continues to serve as a type of ``Ising
model" for chiral symmetry breaking.

\begin{acknowledgments}
We thank Albert Moyano, Joaquim Crusats and Mar\'{i}a-Paz Zorzano
for numerous useful discussions.  One of us (DH) gratefully
acknowledges email correspondence with Thomas Buhse, Jean-Claude
Micheau and Dominique Lavabre.  This research is supported in part
by the Grant AYA2006-15648-C02-01 and -02 from the Ministerio de
Ciencia e Innovaci\'{o}n (Spain), and by the COST Action CM0703
``Systems Chemistry".
\end{acknowledgments}

\end{document}